\documentclass[final,3p,times,twocolumn]{elsarticle}
\usepackage{graphicx,wrapfig,float,slashed,subcaption}
\usepackage{amsmath,amssymb,epsfig,graphicx,graphics,epstopdf}
\usepackage{multirow}
\usepackage{ragged2e}
\usepackage{array}
\usepackage{booktabs} 
\usepackage{relsize}  

\journal{Physics Letters B}

\begin{document}
\begin{frontmatter}

\title 		{Analysis of b quark pair production signal from neutral 2HDM Higgs bosons at future Linear Colliders}
\author[a]	{Majid Hashemi}
\ead			{majid.hashemi@cern.ch}
\author[a]	{Mostafa MahdaviKhorrami}
\ead			{m.mahdavi@shirazu.ac.ir}
\address[a]	{Physics Department, College of Sciences, Shiraz University, Shiraz, 71946-84795, Iran}

\begin{abstract}
In this paper, the b quark pair production events are analyzed as a source of neutral Higgs bosons of the two Higgs doublet model type I at linear colliders. The production mechanism is $e^{+}e^{-} \rightarrow Z^{(*)} \rightarrow HA \rightarrow b\bar{b}b\bar{b}$ assuming a fully hadronic final state. The analysis aim is to identify both CP-even and CP-odd Higgs bosons in different benchmark points accommodating moderate boson masses. Due to pair production of Higgs bosons, the analysis is most suitable for a linear collider operating at $\sqrt{s} = 1$ TeV. Results show that in selected benchmark points, signal peaks are observable in the $b$-jet pair invariant mass distributions at integrated luminosity of 500 $fb^{-1}$.   
\end{abstract}
\end{frontmatter}

\section{Introduction}
One of the most significant accomplishments of standard model (SM) of particle physics is indubitably, observation of Higgs boson at LHC \cite{HiggsObservationCMS,HiggsObservationATLAS} based on a theoretical framework known as the Higgs mechanism  \cite{Englert1,Higgs1,Higgs2,Kibble1,Higgs3,Kibble2}. The observed particle may belong to a single SU(2) doublet (SM Higgs boson) or a model accommodating a larger structure such as two Higgs doublet model (2HDM)  \cite{2hdm1,2hdm2,2hdm3} whose lightest Higgs boson respects the observed particle properties. 

In the latter scenario, one would have a light Higgs boson (h) playing the role of the observed particle, plus additional Higgs bosons with different parities and electric charges. The extra Higgs bosons of the model are assumed to be heavier than the observed one. Therefore, a center of mass energy above the threshold of their masses is required to observe them. Moreover there may be needs for a cleaner collider with a dominant leptonic environment rather than LHC to provide reasonable signature of such particles.

Apart from different scenarios already introduced as beyond standard model (BSM), the 2HDM is considered as a framework for supersymmetry theory, in which each fermion (boson) particle has an associated boson (fermion) particle known as super partner. This theory offers an ingenious solution to the gauge coupling unification at high energies, dark matter candidate (lightest supersymmetric particle) and removal of quadratic divergence of the Higgs boson mass radiative corrections by a natural parameter tuning. In such a theory, the particle space is twice that in SM due to introducing super partners for SM particles and therefore, two Higgs doublets are required to give mass to the double space of particles  \cite{MSSM1,MSSM2,MSSM3}. Here we do not go through such supersymmetric theories, but instead work in the field of 2HDM without supersymmetry.
 
There are four types of 2HDM with different scenarios of Higgs-fermion couplings. The ratio of vacuum expectation values of the two Higgs doublets ($\tan\beta=v_2/v_1$) is the free parameter of the model and is considered as a measure of the Higgs-fermion coupling strength in all 2HDM types \cite{tanbsignificance}.

Our focus in this paper is 2HDM type I which allows for heavy quark pair production in Higgs boson decays while decays to light quarks and leptons are suppressed because the relevant Higgs-fermion couplings depend on the fermion mass. Therefore below the top quark pair production threshold, $H/A \rightarrow b\bar{b}$ is dominant as long as $A \rightarrow Z H$ is not kinematically allowed. 

In other scenarios, such as type III, the Higgs boson coupling with down-type quarks can experience enhancement proportional to $\tan\beta$. However, the current analysis relies on type I in which the signal cross section and Higgs boson decays only depend on the Higgs boson and fermion masses. In recent studies, other types of 2HDM (types II and IV) were analyzed \cite{Hashemi:2014ewa,Hashemi2017,HASHEMI2017_2} leading to overall conclusion that linear colliders have a prominent potential for observation of 2HDM Higgs bosons with a supreme capability over LHC.  

In total five physical Higgs bosons are predicted in 2HDM. The lightest Higgs boson, $h$, (sometimes denoted as $h_{SM}$) is the SM like Higgs boson and there are two heavier neutral Higgs bosons, $H$ (CP-even) and $A$ (CP-odd), and two charged Higgs bosons, $H^{\pm}$. Recently the theory and phenomenology of 2HDM has been extensively discussed in \cite{2hdm_TheoryPheno}.

In addition to direct searches for the 2HDM Higgs bosons which look for direct signals of Higgs boson decays, there are indirect searches based on flavor Physics data. In such searches, deviations from SM observables are looked for when processes containing 2HDM Higgs bosons are added to their corresponding diagrams from SM \cite{FMahmoudi}. Limits obtained from these type of studies have to be taken into account. However, the current analysis focuses on points in parameter space where there is no exclusion from flavor physics studies.

The Higgs boson mass range in this analysis is $150-250$ GeV to be searched for at a future linear collider, operating at $\sqrt{s}=1$ TeV. Signals of heavier Higgs bosons tend to become small when increasing the Higgs bosons masses. All Higgs bosons are assumed to be degenerate in mass, i.e., $m_H=m_A=m_{H^{\pm}}$. This setting ensures that deviation from SM, in terms of $\Delta \rho$, is small enough and consistent with experimental value \cite{drho}. 

The region of interest is $\tan\beta<50$. The signal process, i.e., $e^{+}e^{-} \rightarrow Z^{(*)} \rightarrow HA$ is independent of $\tan\beta$ as the Z-H-A vertex does not depend on Higgs-fermion couplings and the 2HDM type. As will be seen in the next section, the branching ratio of Higgs boson decay to $b\bar{b}$ is also independent of $\tan\beta$ for $\tan\beta < 50$. 

The fully hadronic final state is expected to result in two pairs of $b$-jets (totally four $b$-jets) coming from neutral Higgs boson $H/A$ decays to two $b$-jets. Events which contain four identified (tagged) $b$-jets, are used to produce the $H/A$ invariant mass distribution. The same approach is applied on background events and a final shape discrimination is performed to evaluate the signal significance. Before going to the details of the analysis, a brief review of the theoretical framework is presented in the next section.
\begin{table}
\centering
\begin{tabular}{ccccc}
\midrule
\multicolumn{5}{c}{2HDM Type}\\
\midrule
		 & I & II & III & IV \\
\midrule
$\rho^D$ & $\kappa^D \cot\beta$ &$-\kappa^D \tan\beta$ &$-\kappa^D \tan\beta$ &$\kappa^D \cot\beta$  \\
\midrule
$\rho^U$ & $\kappa^U \cot\beta$ &$\kappa^U \cot\beta$ &$\kappa^U \cot\beta$ &$\kappa^U \cot\beta$  \\
\midrule
$\rho^L$ & $\kappa^L \cot\beta$ &$-\kappa^L \tan\beta$ &$\kappa^L \cot\beta$ &$-\kappa^L \tan\beta$  \\
\midrule
\end{tabular}
\caption{Different types of 2HDM in terms of the Higgs boson couplings with $U$(up-type quarks), $D$(down-type quarks) and $L$(leptons).\label{types}}
\end{table}
\section{Theoretical framework}
Couplings of heavy neutral Higgs bosons ($H$ and $A$) with quarks in Yukawa Lagrangian of the 2HDM, as introduced in \cite{2hdm_HiggsSector1}, takes the form:
\begin{align}
\begin{split}
\mathcal{L}_Y&=\overline{D}\left[\rho^D s_{\beta-\alpha}-\kappa^D c_{\beta-\alpha}\right]DH-i \overline{D}\gamma_{5}\rho^{D}DA\\
&+ \overline{U}\left[\rho^U s_{\beta-\alpha}-\kappa^U c_{\beta-\alpha}\right]UH+i \overline{U}\gamma_{5}\rho^{U}UA
\end{split}
\label{lag1}
\end{align}
in which $U(D)$ are the up(down)-type quarks fields, $H$ and $A$ the neutral Higgs boson fields, $\kappa^q=\frac{m_q}{v}$ for any up(down)-type quark $U(D)$ and $s_{\beta-\alpha}=\sin(\beta-\alpha)$ and $c_{\beta-\alpha}=\cos(\beta-\alpha)$. The $\rho^q$ parameters depend on the 2HDM type and are proportional to $\kappa^q$ as shown in Tab. \ref{types} \cite{Barger_2hdmTypes}. Therefore the four types of interactions (2HDM types) depend on the values of $\rho^f$ which is $\kappa^f$ (SM coupling) times a $\tan\beta$ or $\cot\beta$ factor which makes possible deviations from SM \cite{2hdm_HiggsSector2}.

\begin{figure*}[t!]
    \centering
    \begin{subfigure}[t]{0.45\textwidth}
\centering  \includegraphics[width=\textwidth]{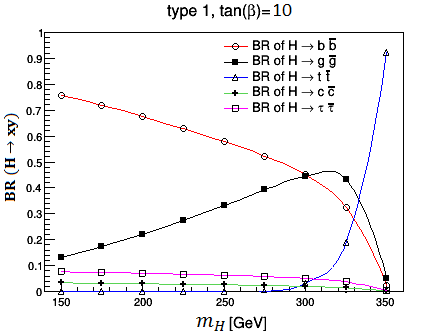}
  \caption{BR(H $\to$ XY) vs $m_H$ \label{BRH_vs_mass}}
    \end{subfigure}%
    ~ 
    \begin{subfigure}[t]{0.45\textwidth}
\centering  \includegraphics[width=\textwidth]{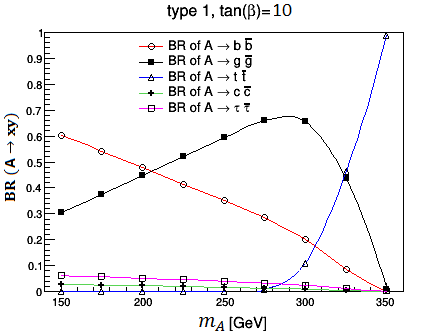}
   \caption{BR(A $\to$ XY) vs $m_A$  \label{BRA_vs_mass}}
	    \end{subfigure}
\caption{The branching ratio of neutral Higgs boson decay in different channels. The $\tan\beta$ is set to 10.}
  \end{figure*}

In Yukawa Lagrangian of the 2HDM type I, the light neutral Higgs ($h$) coupling to fermions and gauge-bosons takes the SM value by setting $s_{\beta-\alpha}=1$. This setting suppresses the heavy neutral Higgs ($H$) coupling with gauge bosons which is proportional to $c_{\alpha-\beta}$\cite{2hdm_TheoryPheno}.
Therefore in our study, we set $s_{\beta-\alpha}=1$, in respect of the correspondence principle so that 2HDM SM-like Higgs boson behaves the same as SM Higgs. This leads to the brief form of the Lagrangian as shown in Eq. \ref{lag1}:
\begin{align}
\begin{split}
\mathcal{L}_Y&=\Bigg\{\overline{D}\rho^DD + \overline{U}\rho^UU\Bigg\}H \\
&-i\Bigg\{\overline{D}\gamma_{5}\rho^{D}D-\overline{U}\gamma_{5}\rho^{U}U\Bigg\}A
\end{split}
\label{lag2}
\end{align}
According to Tab. \ref{types}, the type I appears interesting for low $\tan\beta$ as all couplings in the neutral Higgs sector are proportional to $\cot\beta$. However, the $\cot\beta$ factor cancels out when calculating branching ratio of Higgs boson decays because all Higgs-fermion couplings are proportional to the same factor. This has two subsequences: firstly the Higgs boson decay to fermions is independent of $\tan\beta$ at tree level, and secondly the Higgs boson decays to heavy quarks become dominant due to their larger coupling with the Higgs boson.  As a result, as long as the Higgs boson mass is below the top quark pair production threshold, decay to $b\bar{b}$ is dominant while above the threshold (Higgs boson mass above twice the top quark mass) decay to $t\bar{t}$ starts to grow as seen from Figs. \ref{BRH_vs_mass} and \ref{BRA_vs_mass}. 

\begin{figure}
\centering  \includegraphics[width=0.3\textwidth]{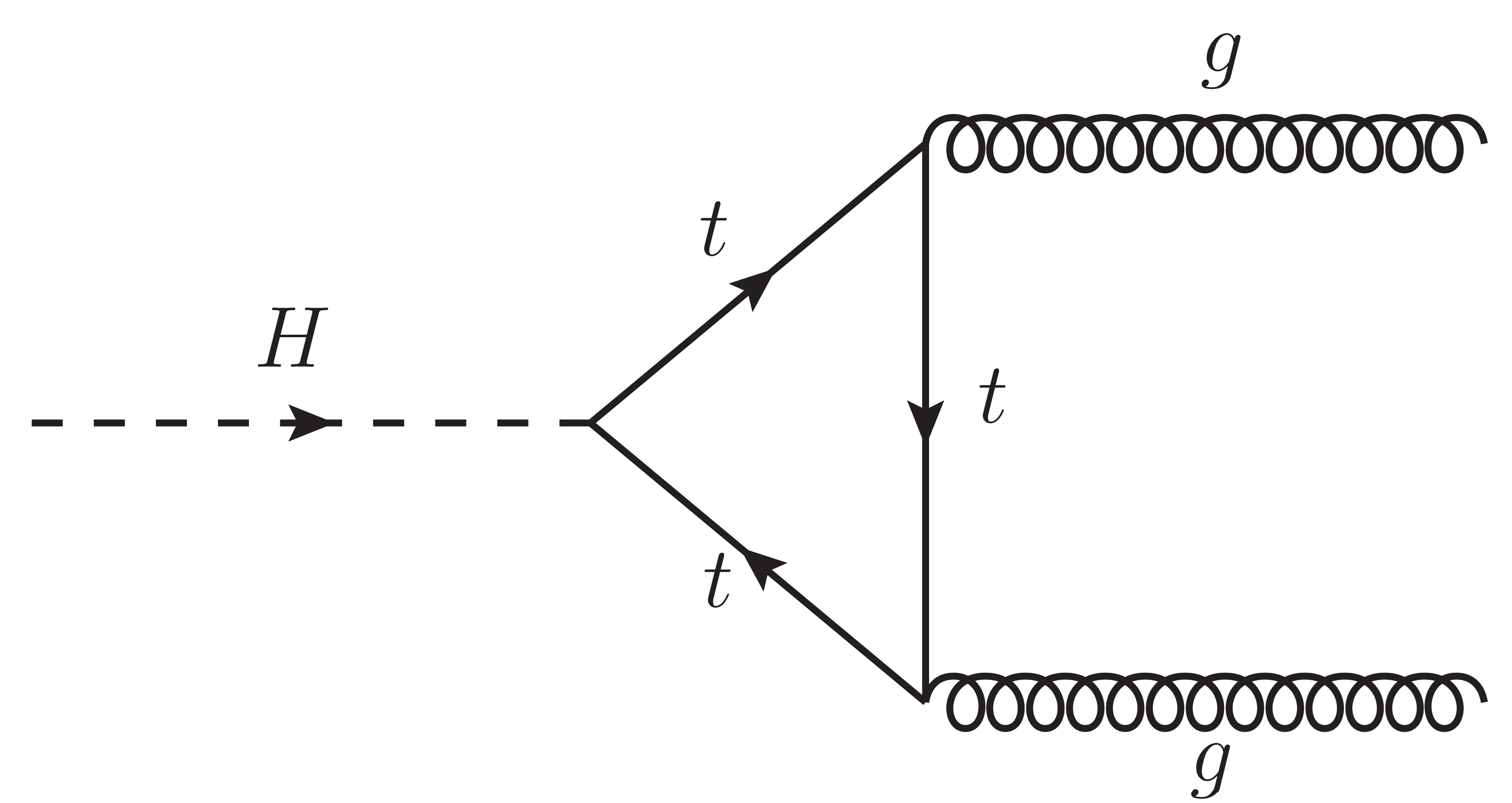}
  \caption{The Feynman diagram of the Higgs boson decay to gluon pair through the virtual top quark loop.}
  \label{H2tt2gg}
\end{figure}

It should be noted that loop diagrams such as $H \rightarrow gg$ proceed through preferably virtual top quarks in the loop, as seen in Figs. \ref{H2tt2gg}. Such decays grow when the Higgs boson mass increases and result in the reduction of decay to $b$ or $c$ quark pairs. The above conclusion is of course valid as long as Higgs boson decay to $t\bar{t}$ is kinematically impossible, i.e.,  $m_{H/A}\lesssim 350$ GeV. 

Although the charged Higgs bosons acquire a strong limit from flavor physics in 2HDM types II and III as reported in \cite{Misiak,misiak2017,FM}, types I and IV receive a small excluded region at low $\tan\beta$. Since the scenario assumed in this paper assumes degenerate Higgs boson states, one may assume the same excluded region for neutral Higgs bosons. However, these limits are at low $\tan\beta$ values and do not affect results of this analysis.

\section{Signal identification and the search scenario}

The signal process, i.e., $e^{+}e^{-} \rightarrow Z^{(*)} \rightarrow HA$ followed by the Higgs bosons decays ($H/A \to b\bar{b}$) depends on the Higgs bosons masses and the cross section decreases when the sum of the masses of the two Higgs bosons reaches the threshold of center of mass energy provided by the collider. As it is shown in Figs. \ref{BRH_vs_mass} and \ref{BRA_vs_mass}, the proper range of the Higgs boson mass is basically $150-350$ GeV. However at masses above 250 GeV, the cross section and decay rates decrease so that a reasonable signal is not observed. The Higgs boson masses are assumed to be equal while a scenario with $m_{A}=m_{H^{\pm}}=m_{H}+50$ GeV is also studied. 

The total cross section of the signal is $\sim12~fb$ corresponding to $m_{H/A} = 150$ GeV and decreases to $\sim9~fb$ with $m_{H}=200$ GeV and $m_{A}=250$ GeV. Taking into account the branching ratio of Higgs bosons decays, these values decrease down to $\sim 5 ~fb$ and $\sim1 ~fb$ respectively. 

All benchmark points are checked to be consistent with the potential stability, perturbativity and unitarity requirements and the current experimental limits on Higgs boson masses using \texttt{2HDMC 1.7.0} \cite{2hdmc1,2hdmc2}. A discussion on this subject will be presented before conclusion.

In order to identify the right pairs of $b$-jets, two methods are tried. The first method is based on the special relativity kinematics; In this method, if the Higgs bosons masses are equal ($m_H = m_A$) which is valid in the scenario presented in this analysis, one can distinguish the two pairs of the $b$-jets. 

\begin{figure*}[]
    \centering
    \begin{subfigure}[t]{0.45\textwidth}
\centering  \includegraphics[width=\textwidth]{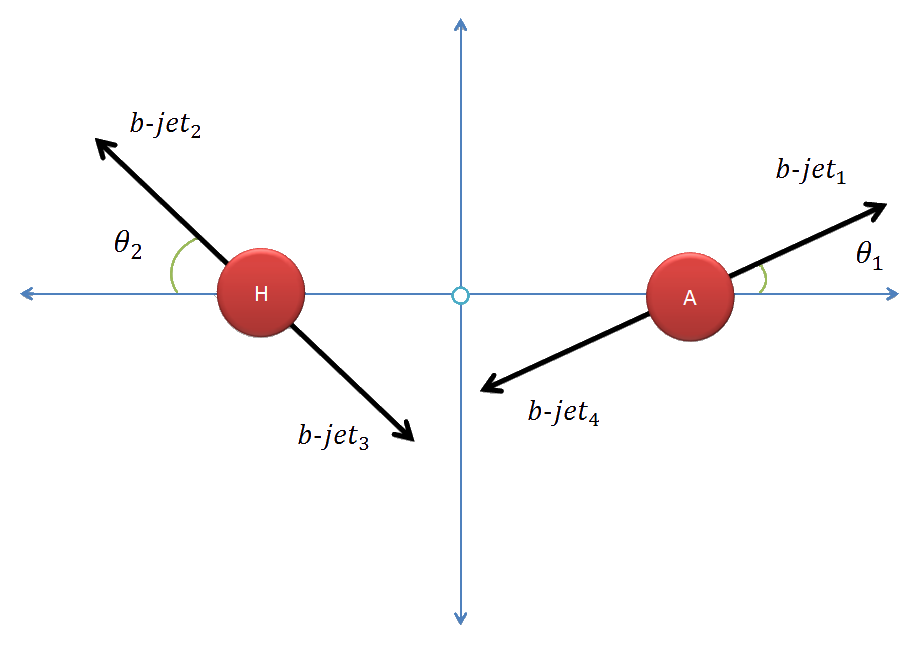}
  \caption{H and A decays as seen in their rest frames }
  \label{HiggsDecay0}
    \end{subfigure}%
    ~ 
    \begin{subfigure}[t]{0.45\textwidth}
\centering  \includegraphics[width=\textwidth]{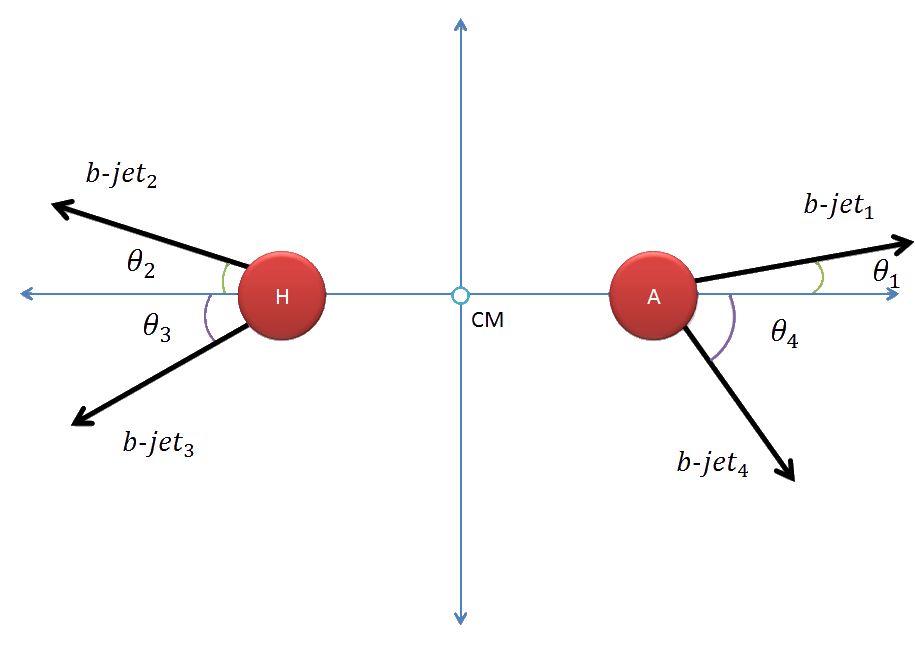}
  \caption{The same decays as seen in the laboratory frame.}
  \label{HiggsDecay}
    \end{subfigure}
\caption{Schematic view of a randomly selected event showing $H$ and $A$ decay to $b\bar{b}$ with each decay picture captured in the rest frame of the decaying particle (a) and in the laboratory frame (b).}
\end{figure*}

Figure \ref{HiggsDecay0} shows a typical event of $H$ and $A$ decays with each decay captured in the rest frame of the decaying Higgs boson. The axis joining the two Higgs bosons can be considered as the preferred axis. The angle between each pair of $b$-jets in the parent Higgs boson rest frame is $\pi$. The $b$-jet flying closest to the axis (most collinear to its parent), has the highest longitudinal momentum component which can be positive or negative when projected to the axis. In the laboratory frame, the event appears in the different form as shown in Fig. \ref{HiggsDecay}. The $b$-jets with smallest angle with respect to the axis in Fig. \ref{HiggsDecay} receive the highest positive or negative Lorentz boost. The two possibilities appear as the maximum and minimum flight angles in the laboratory frame. Due to the momentum conservation perpendicular to the axis, transverse momenta of the two $b$-jets are equal. 

Adding the two longitudinal and transverse momentum components, it is concluded that the $b$-jet with the smallest angle with respect to the axis acquires the maximum total momentum while the one with the largest angle has the minimum total momentum among other $b$-jets. In extreme relativistic limit, the energy and the momentum of a particle are almost equal thus the conclusion holds for particle energies. As the example, according to Fig. \ref{HiggsDecay}, $E_1 > E_4$ and $E_2 > E_3$. Due to the energy conservation, $E_1+E_4=E_2+E_3$, and therefore the $b$-jet with the maximum energy has to come along with the $b$-jet with the lowest energy to keep the sum of the two energies conserved. Without any loss of generality, one may assume $E_1$ to be the highest energy. Then the four $b$-jets energies are sorted like $E_1 > E_2 > E_3 > E_4$ where $E_1,~E_4$ and $E_2,~E_3$ pairs come from their own parents separately. In order to apply this algorithm, the tagged $b$-jets are sorted in terms of their energies and labeled as $b_1,~b_2,~b_3$ and $b_4$. Then two independent pairs, i.e., ($b_1$, $b_4$) and ($b_2$, $b_3$) are used to calculate the invariant masses of the parent Higgs bosons. 

The second method is based on minimizing the angular separation between the two $b$-jets, $\Delta R=\sqrt{(\Delta \eta)^2 + (\Delta \phi)^2 }$, in which $\eta=-\ln(\tan\frac{\theta}{2})$ is the pseudo-rapidity and $\phi$ is the azimuthal angle. This algorithm finds the correct $b$-jet pair with the minimum angular separation. One of the advantages of this method is that it is independent of whether the Higgs bosons masses are equal or not. 

The two methods discussed above were tested separately, however, the second method leads to better results (narrower invariant mass peak, smaller tails and better signal to background ratio due to the less wrong pairing). Therefore results presented in this analysis are based on the second method.
\section{Software setup and cross sections}

In this work, the event generation and cross section calculation of signal processes are handled by \texttt{PYTHIA 8.2.15} \cite{pythia} which uses 2HDM spectrum files in \texttt{LHA} format \cite{lha} generated using \texttt{2HDMC 1.7.0} for each benchmark point separately. The \texttt{LHA} files contain information about the parameters of the model which include Higgs bosons masses, widths and the branching ratio of decay to different channels for each benchmark point. After multi-particle interaction and shower production by PYTHIA, the jet reconstruction is performed by \texttt{FASTJET 3.1} \cite{fastjet1,fastjet2} using anti-$k_t$ algorithm with a jet cone size of 0.4. 

The main SM background processes are $t\bar{t}$, gauge boson pair production $WW$, $ZZ$ and single $Z/\gamma^*$. These background processes are all generated using \texttt{PYTHIA} and their cross sections are also obtained using the same package besides an additional NLO scaling which is applied on the $t\bar{t}$ cross section according to \texttt{MCFM 6.1} \cite{MCFM1,MCFM2,MCFM3,MCFM4}. Tables \ref{sprocess} and \ref{bprocess} indicate the signal benchmark points and background processes and their cross sections.
\begin{table}[h!]
\centering
\begin{tabular}{ccccc}
\midrule
& \multicolumn{4}{c}{\small{Signal Benchmark Points}} \\
\cmidrule(l){2-5}
\small{\small{Higgs Boson Mass}}		& \small{BP1} & \small{BP2} & \small{BP3} & \small{BP4}\\
\midrule
$m_h$									& \multicolumn{4}{c}{\small{125}} \\
\midrule
$m_H$									& \small{150} & \small{200} & \small{250} & \small{200} \\
\midrule
$m_A$									& \small{150} & \small{200} & \small{250} & \small{250} \\
\midrule
$\tan\beta$							& \multicolumn{4}{c}{\small{10}}\\
\midrule
\small{Cross Section $[fb]$}		&\small{12.2} & \small{10.4} & \small{8.45} & \small{9.36} \\
\midrule
\small{BR$_{totall}(=BR_H*BR_A$) $\%$}	&\small{45.6}&\small{32.2}&\small{20.2}&\small{11.1} \\
\midrule
\small{Cross Section $*$ BR$_{totall} [fb]$}	&\small{5.56}&\small{3.35}&\small{1.71}&\small{1.04} \\
\bottomrule
\end{tabular}
\caption{Signal benchmark points and their cross sections and branching ratios used in the analysis. BR denotes the branching ratio of Higgs boson decay to $b\bar{b}$.}
\label{sprocess}
\begin{tabular}{ccccc}
\\
\\
\midrule
& \multicolumn{4}{c}{\small{Backgrounds}}\\
\cmidrule(l){2-5}
\small{Channel}&\small{$Z/\gamma*$}&\small{$ZZ$}&\small{$WW$}&\small{$t\bar{t}$}\\
\midrule
\small{Cross Sections $[fb]$}  & \small{547} & \small{4.54} & \small{1460} & \small{96.4} \\
\bottomrule
\end{tabular}
\caption {Background processes and their cross sections. In $Z/\gamma^*$ and $ZZ$ events, only the $Z$ boson decay to $b\bar{b}$ is considered, while for $WW$ and $t\bar{t}$, the fully hadronic decay is produced by forcing the W boson to decay to all quark pairs.}
\label{bprocess}
\end{table}
\section{Signal selection and analysis}
The event selection starts from $b$-tagging using reconstructed jets from \texttt{FASTJET}. A kinematic requirement is applied on jets as $E_j > 5. ~\textnormal{GeV},~~~|\eta|<5.$ to reject soft jets or jets close to the beam pipe. The $b$-tagging algorithm is based on a simplified matching between the reconstructed jet and the $b$-quark in the event. For each reconstructed jet, a search for $b$-quarks in the jet cone is performed. If the jet accommodates a $b$ or $c$ quark in the reconstruction cone, it is tagged as a $b$-jet with a $b$-tagging probability of 70$\%$ and fake rate of 10$\%$ (from c-jets).\\ Each event has to have four $b$-jets to be analyzed. A search among the four $b$-jets in the event is performed to find the correct $b$-jet pair with minimum $\Delta R$ and calculates their invariant mass. The same algorithm is applied on background processes. The distributions of $b$-jet pair invariant masses are stored in histograms for both signal and background processes. 

\begin{figure*}[h]
    \centering
    \begin{subfigure}[h]{0.45\textwidth}
        \centering
        \includegraphics[width=\textwidth]{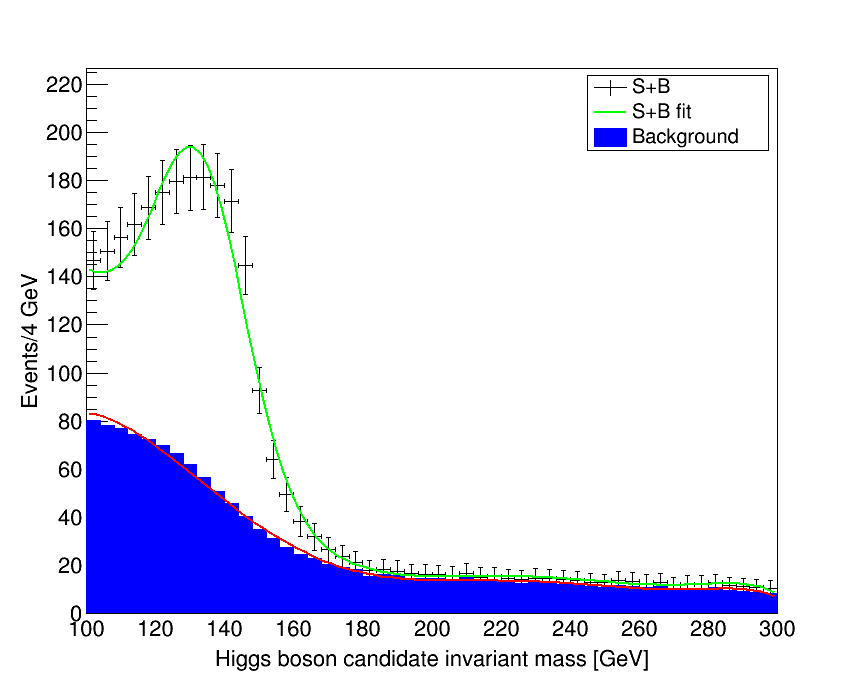}
        \caption{$m_H=m_A=150$ GeV}
\label{s1}
    \end{subfigure}%
    ~ 
    \begin{subfigure}[h]{0.45\textwidth}
        \centering
        \includegraphics[width=\textwidth]{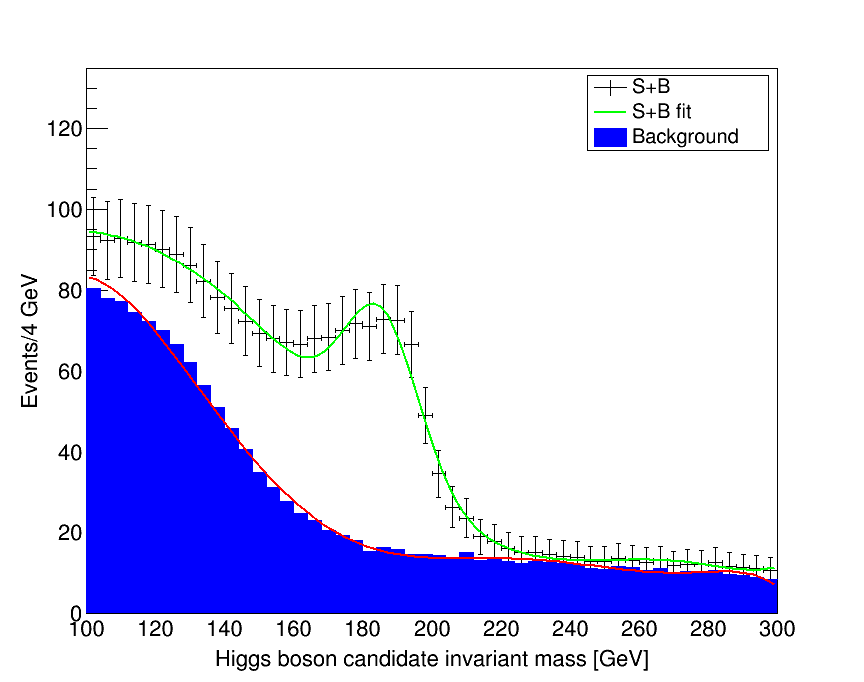}
        \caption{$m_H=m_A=200$ GeV}
\label{s2}
    \end{subfigure}
    ~ 
    \begin{subfigure}[h]{0.45\textwidth}
        \centering
        \includegraphics[width=\textwidth]{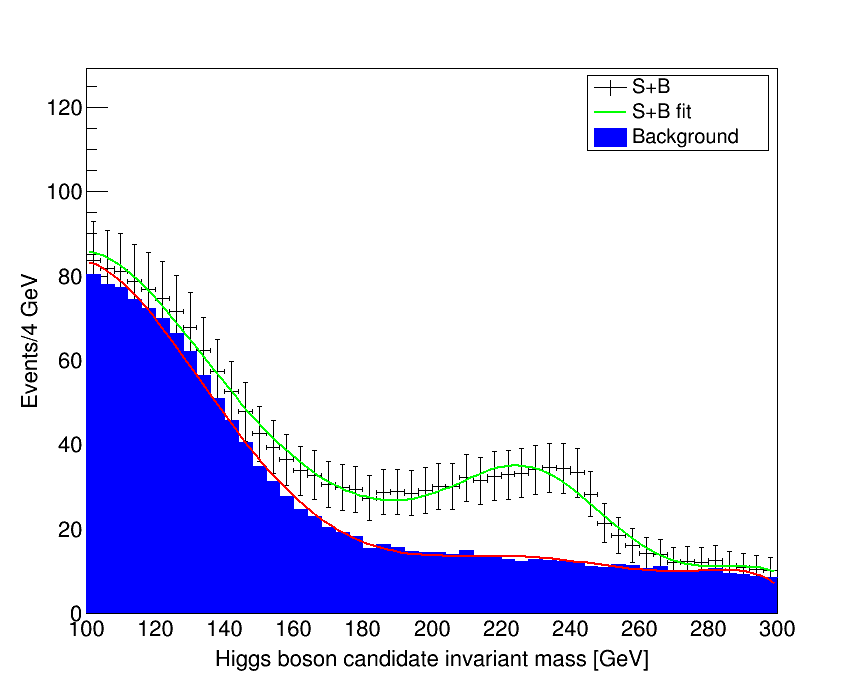}
        \caption{$m_H=m_A=250$ GeV}
  \label{s3}
  \end{subfigure}
    ~ 
    \begin{subfigure}[h]{0.45\textwidth}
        \centering
        \includegraphics[width=\textwidth]{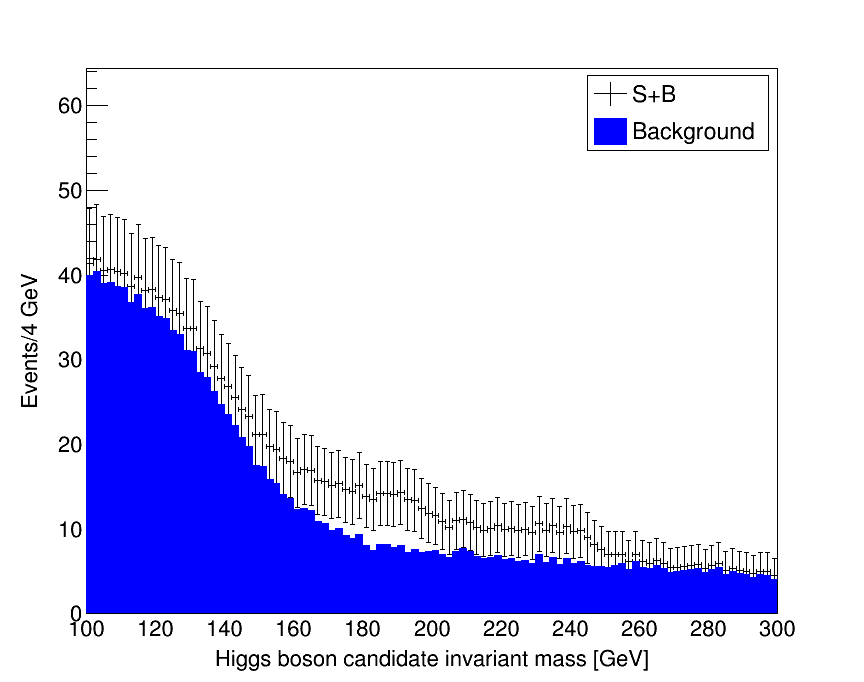}
        \caption{$m_H=200$ GeV, $m_A=250$ GeV}
  \label{s4}
  \end{subfigure}
    \caption{The $b$-jet pair invariant mass in signal and background processes.}
\end{figure*}

In order to select events in the signal region (region of the signal peak in the $b$-jet pair invariant mass distribution), a mass window is applied on the signal plus background distributions. The mass window enhances the signal to background ratio leading to a higher signal purity. Figures \ref{s1}, \ref{s2}, \ref{s3} show the distributions of $b\bar{b}$ invariant mass from signal events on top the background. 

A polynomial fit describes the background distribution and is used as the input probability distribution function (PDF) for the background (the red curves shown in Figs. \ref{s1}, \ref{s2}, \ref{s3}). A Gaussian fit is then added to the background PDF (the polynomial function) and is applied on signal plus background distribution using the input parameter values taken from the previous step. The result of the fits are shown as green curves in Figs. \ref{s1}, \ref{s2}, \ref{s3}. The error bars are taken into account in the fit which is based on $\chi^2$ minimization. They are however of statistical origin and systematic uncertainties are not taken into account. 

Table \ref{effs} shows mass windows and the total signal and background efficiencies and the final number of signal and background. The signal to background ratio and signal statistical significance in different points are also included. The mass window is set by maximizing the signal significance in a search for the best position of the left and right sides of the window. As seen from Tab. \ref{effs}, the signal significance is reasonable and above $5\sigma$ for all selected benchmark points.

It should be noted that we obtain reasonable results only for BP1, BP2 and BP3. The fourth benchmark (BP4) which involves different Higgs boson masses leads to smaller branching ratio of Higgs boson decays compared to other points. The reason is mainly due to the CP-odd Higgs decay to ZH $(A \to ZH)$ which turns on when there is enough phase space for the decay even through off-shell production of the decay products. We examined this scenario and the result is shown in Fig. \ref{s4}. The signal significance in this case is $~7\sigma$. However, fitting this distribution by including two Gaussian fits on top of a polinomial is challenging due to the small size of the signal and large error bars.    

\begin{table*}[h]
\centering
\begin{tabular}{ccccccccccc}
\midrule
\small{Benchmark points} &&& BP1 &&& BP2 &&& BP3 \\
\midrule
\small{$m_H=m_A$ $[GeV]$}	&&& 150 &&&  200 &&& 250 \\
\midrule
\small{Mass Window [GeV]}			&&& 112-152 &&& 160-200 &&& 208-248 &\\
\midrule
\small{Int. luminosity [$fb^{-1}$]}		&&& 500 &&& 500 &&& 500 &  \\
\midrule
\small{Total efficiency (S)}	&&& 0.7081 &&& 0.7243 &&& 0.7287 &  \\
\midrule
\small{Total efficiency (B)}	&&& 0.2474 &&& 0.2474 &&& 0.2474 &  \\
\midrule
\small{$S$}							&&& 1140 &&& 533.1 &&& 214.8 &  \\
\midrule
\small{$B$}							&&& 651.3 &&& 210.8 &&& 142.7 &  \\
\midrule
$\frac{S}{B}$						&&& 1.750 &&& 2.528 &&& 1.506 &  \\
\midrule
$\frac{S}{\sqrt{B}}$					&&& 44.66 &&& 36.71 &&& 17.98 &  \\
\midrule
\end{tabular}
\caption {($\mathlarger{\mathlarger m_H=m_A=m_{H^{\pm}}}$)}
\label{effs}
\end{table*}
\section{Discussion}
\begin{figure}[h]
        \centering
        \includegraphics[width=0.5\textwidth,height=0.4\textwidth]{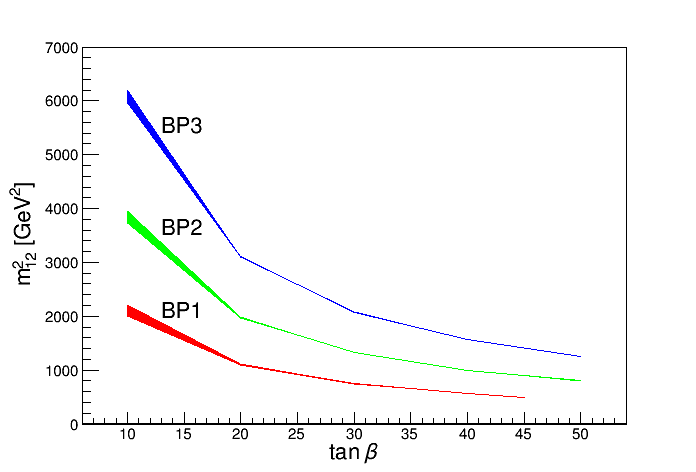}
        \caption{The allowed $m_{12}^2$ values satisfying theoretical requirements as a function of $\tan\beta$ for different benchmark points.}
  \label{m12}
  \end{figure}

It should be noted that the theoretical space available to this study is limited in $m_{12}^2$ vs $\tan\beta$ parameter space. In order to respect theoretical requirements, a scan over possible $m_{12}^2$ values was performed for each $\tan\beta$ value resulting in a narrow range of $m_{12}^2$ which satisfies all requirements of potential stability, perturbativity and unitarity. Results are plotted in Fig. \ref{m12}. As seen from Fig. \ref{m12}, the available range of $m_{12}^2$ becomes smaller with increasing $\tan\beta$ for each benchmark point. Any value of $m_{12}^2$ in the allowed range leads to the same branching ratio of Higgs boson decay to $b\bar{b}$ for $\tan\beta<50$. Therefore the signal significance values obtained for $\tan\beta=10$ are valid up to $\tan\beta\simeq 50$ provided that for each value of $\tan\beta$, the allowed value of $m_{12}^2$ is taken from the range provided in Fig. \ref{m12}.
\section{Conclusions}
The observability of 2HDM Higgs bosons was studied at a $e^+e^-$ linear collider operating at $\sqrt{s}=1$ TeV. Different benchmark points were studied focusing on moderate values of the Higgs bosons masses. The theoretical framework was set to 2HDM type I where the Higgs (H or A)  boson decay to $b\bar{b}$ is dominant as long as the Higgs boson mass is below the threshold of on-shell decay to $t\bar{t}$. The leptonic decays are also suppressed as $\cot\beta$. The signal cross section is almost independent of $\tan\beta$ and therefore all results were quoted based on a typical $\tan\beta$ value of 10. The collider integrated luminosity was set to 500 $fb^{-1}$. Results show that the Higgs boson signal in $b\bar{b}$ invariant mass distribution is well observable for all benchmark points. There is also a fit possibility for those points with equal Higgs boson masses.  
\section*{Acknowledgments}
We would like to thank the college of sciences at Shiraz university for providing computational facilities and maintaining the computing cluster during the research program.
\section*{References}
\bibliographystyle{elsarticle-num}
\bibliography{BIB_TO_USE}

\begin{thebibliography}{10}
\expandafter\ifx\csname url\endcsname\relax
  \def\url#1{\texttt{#1}}\fi
\expandafter\ifx\csname urlprefix\endcsname\relax\def\urlprefix{URL }\fi
\expandafter\ifx\csname href\endcsname\relax
  \def\href#1#2{#2} \def\path#1{#1}\fi

\bibitem{HiggsObservationCMS}
S.~Chatrchyan, et~al., {Observation of a new boson at a mass of 125 GeV with
  the CMS experiment at the LHC}, Phys. Lett. B716 (2012) 30--61.
\newblock \href {http://arxiv.org/abs/1207.7235} {\path{arXiv:1207.7235}},
  \href {http://dx.doi.org/10.1016/j.physletb.2012.08.021}
  {\path{doi:10.1016/j.physletb.2012.08.021}}.

\bibitem{HiggsObservationATLAS}
G.~Aad, et~al., {Observation of a new particle in the search for the Standard
  Model Higgs boson with the ATLAS detector at the LHC}, Phys. Lett. B716
  (2012) 1--29.
\newblock \href {http://arxiv.org/abs/1207.7214} {\path{arXiv:1207.7214}},
  \href {http://dx.doi.org/10.1016/j.physletb.2012.08.020}
  {\path{doi:10.1016/j.physletb.2012.08.020}}.

\bibitem{Englert1}
F.~Englert, R.~Brout, {Broken Symmetry and the Mass of Gauge Vector Mesons},
  Phys. Rev. Lett. 13 (1964) 321--323.
\newblock \href {http://dx.doi.org/10.1103/PhysRevLett.13.321}
  {\path{doi:10.1103/PhysRevLett.13.321}}.

\bibitem{Higgs1}
P.~W. Higgs, {Broken Symmetries and the Masses of Gauge Bosons}, Phys. Rev.
  Lett. 13 (1964) 508--509.
\newblock \href {http://dx.doi.org/10.1103/PhysRevLett.13.508}
  {\path{doi:10.1103/PhysRevLett.13.508}}.

\bibitem{Higgs2}
P.~W. Higgs, {Broken symmetries, massless particles and gauge fields}, Phys.
  Lett. 12 (1964) 132--133.
\newblock \href {http://dx.doi.org/10.1016/0031-9163(64)91136-9}
  {\path{doi:10.1016/0031-9163(64)91136-9}}.

\bibitem{Kibble1}
G.~S. Guralnik, C.~R. Hagen, T.~W.~B. Kibble, {Global Conservation Laws and
  Massless Particles}, Phys. Rev. Lett. 13 (1964) 585--587.
\newblock \href {http://dx.doi.org/10.1103/PhysRevLett.13.585}
  {\path{doi:10.1103/PhysRevLett.13.585}}.

\bibitem{Higgs3}
P.~W. Higgs, {Spontaneous Symmetry Breakdown without Massless Bosons}, Phys.
  Rev. 145 (1966) 1156--1163.
\newblock \href {http://dx.doi.org/10.1103/PhysRev.145.1156}
  {\path{doi:10.1103/PhysRev.145.1156}}.

\bibitem{Kibble2}
T.~W.~B. Kibble, {Symmetry breaking in nonAbelian gauge theories}, Phys. Rev.
  155 (1967) 1554--1561.
\newblock \href {http://dx.doi.org/10.1103/PhysRev.155.1554}
  {\path{doi:10.1103/PhysRev.155.1554}}.

\bibitem{2hdm1}
T.~D. Lee, {A Theory of Spontaneous T Violation}, Phys. Rev. D8 (1973)
  1226--1239.
\newblock \href {http://dx.doi.org/10.1103/PhysRevD.8.1226}
  {\path{doi:10.1103/PhysRevD.8.1226}}.

\bibitem{2hdm2}
S.~L. Glashow, S.~Weinberg, {Natural Conservation Laws for Neutral Currents},
  Phys. Rev. D15 (1977) 1958.
\newblock \href {http://dx.doi.org/10.1103/PhysRevD.15.1958}
  {\path{doi:10.1103/PhysRevD.15.1958}}.

\bibitem{2hdm3}
G.~C. Branco, {Spontaneous {CP} Nonconservation and Natural Flavor
  Conservation: A Minimal Model}, Phys. Rev. D22 (1980) 2901.
\newblock \href {http://dx.doi.org/10.1103/PhysRevD.22.2901}
  {\path{doi:10.1103/PhysRevD.22.2901}}.

\bibitem{MSSM1}
I.~J.~R. Aitchison, {Supersymmetry and the MSSM: An Elementary
  introduction}\href {http://arxiv.org/abs/hep-ph/0505105}
  {\path{arXiv:hep-ph/0505105}}.

\bibitem{MSSM2}
E.~Ma, D.~Ng, {New supersymmetric option for two Higgs doublets}, Phys. Rev.
  D49 (1994) 6164--6167.
\newblock \href {http://arxiv.org/abs/hep-ph/9305230}
  {\path{arXiv:hep-ph/9305230}}, \href
  {http://dx.doi.org/10.1103/PhysRevD.49.6164}
  {\path{doi:10.1103/PhysRevD.49.6164}}.

\bibitem{MSSM3}
A.~Djouadi, {The Anatomy of electro-weak symmetry breaking. II. The Higgs
  bosons in the minimal supersymmetric model}, Phys. Rept. 459 (2008) 1--241.
\newblock \href {http://arxiv.org/abs/hep-ph/0503173}
  {\path{arXiv:hep-ph/0503173}}, \href
  {http://dx.doi.org/10.1016/j.physrep.2007.10.005}
  {\path{doi:10.1016/j.physrep.2007.10.005}}.

\bibitem{tanbsignificance}
H.~E. Haber, D.~O'Neil, {Basis-independent methods for the two-Higgs-doublet
  model. II. The Significance of tan$\beta$}, Phys. Rev. D74 (2006) 015018,
  [Erratum: Phys. Rev.D74,no.5,059905(2006)].
\newblock \href {http://arxiv.org/abs/hep-ph/0602242}
  {\path{arXiv:hep-ph/0602242}}, \href
  {http://dx.doi.org/10.1103/PhysRevD.74.015018, 10.1103/PhysRevD.74.059905}
  {\path{doi:10.1103/PhysRevD.74.015018, 10.1103/PhysRevD.74.059905}}.

\bibitem{Hashemi:2014ewa}
M.~Hashemi, I.~Ahmed, {Observability of triple or double charged Higgs
  production in two Higgs doublet model type II at an e + e - linear collider},
  Int. J. Mod. Phys. A30~(04n05) (2015) 1550022.
\newblock \href {http://arxiv.org/abs/1401.4841} {\path{arXiv:1401.4841}},
  \href {http://dx.doi.org/10.1142/S0217751X15500220}
  {\path{doi:10.1142/S0217751X15500220}}.

\bibitem{Hashemi2017}
M.~Hashemi, Leptophilic neutral higgs bosons in two higgs doublet model at a
  linear collider, The European Physical Journal C 77 (2017) 302.
\newblock \href {http://dx.doi.org/10.1140/epjc/s10052-017-4863-0}
  {\path{doi:10.1140/epjc/s10052-017-4863-0}}.

\bibitem{HASHEMI2017_2}
M.~Hashemi, G.~Haghighat, Search for heavy neutral cp-even higgs within
  lepton-specific 2hdm at a future linear collider, Physics Letters B 772
  (2017) 426 -- 434.
\newblock \href {http://dx.doi.org/10.1016/j.physletb.2017.06.068}
  {\path{doi:10.1016/j.physletb.2017.06.068}}.

\bibitem{2hdm_TheoryPheno}
G.~C. Branco, P.~M. Ferreira, L.~Lavoura, M.~N. Rebelo, M.~Sher, J.~P. Silva,
  {Theory and phenomenology of two-Higgs-doublet models}, Phys. Rept. 516
  (2012) 1--102.
\newblock \href {http://arxiv.org/abs/1106.0034} {\path{arXiv:1106.0034}},
  \href {http://dx.doi.org/10.1016/j.physrep.2012.02.002}
  {\path{doi:10.1016/j.physrep.2012.02.002}}.

\bibitem{FMahmoudi}
F.~Mahmoudi, O.~Stal, {Flavor constraints on the two-Higgs-doublet model with
  general Yukawa couplings}, Phys. Rev. D81 (2010) 035016.
\newblock \href {http://arxiv.org/abs/0907.1791} {\path{arXiv:0907.1791}},
  \href {http://dx.doi.org/10.1103/PhysRevD.81.035016}
  {\path{doi:10.1103/PhysRevD.81.035016}}.

\bibitem{drho}
W.~Grimus, L.~Lavoura, O.~M. Ogreid, P.~Osland, {A Precision constraint on
  multi-Higgs-doublet models}, J. Phys. G35 (2008) 075001.
\newblock \href {http://arxiv.org/abs/0711.4022} {\path{arXiv:0711.4022}},
  \href {http://dx.doi.org/10.1088/0954-3899/35/7/075001}
  {\path{doi:10.1088/0954-3899/35/7/075001}}.

\bibitem{2hdm_HiggsSector1}
S.~Davidson, H.~E. Haber, {Basis-independent methods for the two-Higgs-doublet
  model}, Phys. Rev. D72 (2005) 035004, [Erratum: Phys. Rev.D72,099902(2005)].
\newblock \href {http://arxiv.org/abs/hep-ph/0504050}
  {\path{arXiv:hep-ph/0504050}}, \href
  {http://dx.doi.org/10.1103/PhysRevD.72.099902, 10.1103/PhysRevD.72.035004}
  {\path{doi:10.1103/PhysRevD.72.099902, 10.1103/PhysRevD.72.035004}}.

\bibitem{Barger_2hdmTypes}
V.~D. Barger, J.~L. Hewett, R.~J.~N. Phillips, {New Constraints on the Charged
  Higgs Sector in Two Higgs Doublet Models}, Phys. Rev. D41 (1990) 3421--3441.
\newblock \href {http://dx.doi.org/10.1103/PhysRevD.41.3421}
  {\path{doi:10.1103/PhysRevD.41.3421}}.

\bibitem{2hdm_HiggsSector2}
M.~Aoki, S.~Kanemura, K.~Tsumura, K.~Yagyu, {Models of Yukawa interaction in
  the two Higgs doublet model, and their collider phenomenology}, Phys. Rev.
  D80 (2009) 015017.
\newblock \href {http://arxiv.org/abs/0902.4665} {\path{arXiv:0902.4665}},
  \href {http://dx.doi.org/10.1103/PhysRevD.80.015017}
  {\path{doi:10.1103/PhysRevD.80.015017}}.

\bibitem{Misiak}
M.~Misiak, et~al., {Updated NNLO QCD predictions for the weak radiative B-meson
  decays}, Phys. Rev. Lett. 114~(22) (2015) 221801.
\newblock \href {http://arxiv.org/abs/1503.01789} {\path{arXiv:1503.01789}},
  \href {http://dx.doi.org/10.1103/PhysRevLett.114.221801}
  {\path{doi:10.1103/PhysRevLett.114.221801}}.

\bibitem{misiak2017}
M.~Misiak, M.~Steinhauser, Weak radiative decays of the b meson and bounds on
  $m_{H^{\pm}}$ in the two-higgs-doublet model, The European Physical Journal C
  77 (2017) 201.
\newblock \href {http://dx.doi.org/10.1140/epjc/s10052-017-4776-y}
  {\path{doi:10.1140/epjc/s10052-017-4776-y}}.

\bibitem{FM}
A.~Arbey, F.~Mahmoudi, O.~Stal, T.~Stefaniak, {Status of the Charged Higgs
  Boson in Two Higgs Doublet Models. }\href {http://arxiv.org/abs/1706.07414}
  {\path{arXiv:1706.07414}}.

\bibitem{2hdmc1}
D.~Eriksson, J.~Rathsman, O.~Stal, {2HDMC: Two-Higgs-Doublet Model Calculator
  Physics and Manual}, Comput. Phys. Commun. 181 (2010) 189--205.
\newblock \href {http://arxiv.org/abs/0902.0851} {\path{arXiv:0902.0851}},
  \href {http://dx.doi.org/10.1016/j.cpc.2009.09.011}
  {\path{doi:10.1016/j.cpc.2009.09.011}}.

\bibitem{2hdmc2}
D.~Eriksson, J.~Rathsman, O.~Stal, {2HDMC: Two-Higgs-doublet model calculator},
  Comput. Phys. Commun. 181 (2010) 833--834.
\newblock \href {http://dx.doi.org/10.1016/j.cpc.2009.12.016}
  {\path{doi:10.1016/j.cpc.2009.12.016}}.

\bibitem{pythia}
T.~Sjostrand, S.~Mrenna, P.~Z. Skands, {A Brief Introduction to PYTHIA 8.1},
  Comput. Phys. Commun. 178 (2008) 852--867.
\newblock \href {http://arxiv.org/abs/0710.3820} {\path{arXiv:0710.3820}},
  \href {http://dx.doi.org/10.1016/j.cpc.2008.01.036}
  {\path{doi:10.1016/j.cpc.2008.01.036}}.

\bibitem{lha}
J.~Alwall, et~al., {A Standard format for Les Houches event files}, Comput.
  Phys. Commun. 176 (2007) 300--304.
\newblock \href {http://arxiv.org/abs/hep-ph/0609017}
  {\path{arXiv:hep-ph/0609017}}, \href
  {http://dx.doi.org/10.1016/j.cpc.2006.11.010}
  {\path{doi:10.1016/j.cpc.2006.11.010}}.

\bibitem{fastjet1}
M.~Cacciari, {FastJet: A Code for fast $k_t$ clustering, and more}, in: {Deep
  inelastic scattering. Proceedings, 14th International Workshop, DIS 2006,
  Tsukuba, Japan, April 20-24, 2006}, 2006, pp. 487--490, [,125(2006)].
\newblock \href {http://arxiv.org/abs/hep-ph/0607071}
  {\path{arXiv:hep-ph/0607071}}.

\bibitem{fastjet2}
M.~Cacciari, G.~P. Salam, G.~Soyez, {FastJet User Manual}, Eur. Phys. J. C72
  (2012) 1896.
\newblock \href {http://arxiv.org/abs/1111.6097} {\path{arXiv:1111.6097}},
  \href {http://dx.doi.org/10.1140/epjc/s10052-012-1896-2}
  {\path{doi:10.1140/epjc/s10052-012-1896-2}}.

\bibitem{MCFM1}
J.~M. Campbell, R.~K. Ellis, F.~Tramontano, {Single top production and decay at
  next-to-leading order}, Phys. Rev. D70 (2004) 094012.
\newblock \href {http://arxiv.org/abs/hep-ph/0408158}
  {\path{arXiv:hep-ph/0408158}}, \href
  {http://dx.doi.org/10.1103/PhysRevD.70.094012}
  {\path{doi:10.1103/PhysRevD.70.094012}}.

\bibitem{MCFM2}
J.~M. Campbell, F.~Tramontano, {Next-to-leading order corrections to Wt
  production and decay}, Nucl. Phys. B726 (2005) 109--130.
\newblock \href {http://arxiv.org/abs/hep-ph/0506289}
  {\path{arXiv:hep-ph/0506289}}, \href
  {http://dx.doi.org/10.1016/j.nuclphysb.2005.08.015}
  {\path{doi:10.1016/j.nuclphysb.2005.08.015}}.

\bibitem{MCFM3}
J.~M. Campbell, R.~Frederix, F.~Maltoni, F.~Tramontano, {Next-to-Leading-Order
  Predictions for t-Channel Single-Top Production at Hadron Colliders}, Phys.
  Rev. Lett. 102 (2009) 182003.
\newblock \href {http://arxiv.org/abs/0903.0005} {\path{arXiv:0903.0005}},
  \href {http://dx.doi.org/10.1103/PhysRevLett.102.182003}
  {\path{doi:10.1103/PhysRevLett.102.182003}}.

\bibitem{MCFM4}
J.~M. Campbell, R.~K. Ellis, {Top-quark processes at NLO in production and
  decay}, J. Phys. G42~(1) (2015) 015005.
\newblock \href {http://arxiv.org/abs/1204.1513} {\path{arXiv:1204.1513}},
  \href {http://dx.doi.org/10.1088/0954-3899/42/1/015005}
  {\path{doi:10.1088/0954-3899/42/1/015005}}.

\end{thebibliography}
\end{document}